\documentclass[11pt]{article}
\usepackage[dvips]{graphics}
\title{Many-fingered time Bohmian mechanics}
\author{Hrvoje Nikoli\'c \\
Theoretical Physics Division, Rudjer Bo\v{s}kovi\'{c} Institute, \\
P.O.B. 180, HR-10002 Zagreb, Croatia \\
{\normalsize e-mail: hrvoje@thphys.irb.hr} \\
\makebox[1in]{} \\
}
\date{\today}
\begin{document}
\maketitle
\begin{abstract}
The many-fingered time (MFT) formulation of many-particle quantum mechanics 
and quantum field theory is a natural framework that overcomes 
the problem of ``instantaneous collapse" in entangled systems that 
exhibit nonlocalities.
The corresponding Bohmian interpretation can also be formulated 
in terms of MFT beables, which alleviates the problem of 
instantaneous action at a distance by using an ontology that differs 
from that in the standard Bohmian interpretation. 
The appearance of usual single-time 
particle-positions and fields is recovered by quantum measurements.    
\end{abstract}

\vspace*{0.5cm}


\maketitle

\section{Introduction}

Entanglement in quantum mechanics (QM) induces certain nonlocal features 
of QM. While there is still some controversy regarding the question
if orthodox QM itself is really an intrinsically nonlocal theory
(see e.g. \cite{medina} and references therein), from the work 
of John Bell \cite{bell} it is clear that any hidden-variable 
interpretation of QM must be explicitly nonlocal. The best known 
and most successful nonlocal hidden-variable interpretation of QM
and quantum field theory (QFT) is the Bohmian interpretation
\cite{bohm12,bohmrep1,bohmrep2,holrep,holbook,nikfpl12}. 
A typical property of this interpretation is an    
instantaneous action at a distance among the hidden-variables -- 
particle-positions and field-configurations. The word
``instantaneous" requires a preferred global choice of the 
time-coordinate, which seems to contradict the principle of
relativity. A possible way out of this problem is to 
introduce a ``preferred" foliation of spacetime in a 
dynamical way \cite{durr99,hort,nikolDDW}. Another possibility 
is to introduce a Bohmian equation of motion not only for 
space-coordinates of particles, but also for their 
time-coordinates \cite{durr96,nikolcov}. 

The most recent possibility, 
suggested in \cite{nikolpla} for quantum fields,
is the {\em many-fingered time} (MFT) 
formulation of Bohmian mechanics, based on the MFT formulation 
of orthodox many-particle QM \cite{tomo} and QFT \cite{tomo,schw}.
The purpose of the present paper is to further develop the idea 
of the MFT Bohmian interpretation introduced in \cite{nikolpla}.
More specifically, the aim is (i) to present the MFT formulation 
of Bohmian mechanics for many-particle QM (which was not presented 
in \cite{nikolpla})
and (ii) to improve and correct some of the results and statements 
on the MFT Bohmian mechanics of fields presented in \cite{nikolpla}.
The present paper can also be viewed as complementary to \cite{nikolpla},
in the sense that the present paper, unlike 
\cite{nikolpla}, does not insist on the 
manifestly relativistic-covariant formulation, 
but instead discusses the conceptual meaning 
of the MFT-nature of Bohmian hidden-variable beables more carefully.

Sec.~\ref{MFT} contains a review of the orthodox MFT formulation of
many-particle QM, while the corresponding MFT Bohmian interpretation is 
discussed in Sec.~\ref{BOHM}. The generalization to QFT is briefly 
discussed in Sec.~\ref{QFT}, after which the conclusions are drawn in 
Sec.~\ref{CONCL}.  

Throughout the paper, we use units in which $\hbar=1$.

\section{MFT formulation of many-particle QM} 
\label{MFT}      

A natural starting point towards a relativistic-covariant formulation 
of many-particle QM is to introduce a kinematical 
framework in which time is treated on an equal footing with space.
Thus, instead of a single-time $n$-particle wave function
$\psi({\bf x}_1,\ldots,{\bf x}_n,t)$, one introduces a 
MFT $n$-particle wave function \cite{tomo}
\begin{equation}
\Psi({\bf x}_1,\ldots,{\bf x}_n,t_1,\ldots,t_n).
\end{equation} 
However, a MFT formulation can also be introduced independently 
of the principle of relativity, so in this section,
for simplicity, we actually 
study the nonrelativistic version of the MFT formulation of QM.
One of the 
main purposes of this study is to demonstrate that, with the MFT
formulation of QM, the wave-function ``collapse" induced 
by a measurement does not require a preferred notion of 
simultaneity. 

The quantity 
\begin{equation}
\rho({\bf x}_1,\ldots,{\bf x}_n,t_1,\ldots,t_n)=
|\Psi({\bf x}_1,\ldots,{\bf x}_n,t_1,\ldots,t_n)|^2 
\end{equation}
is the probability density for finding one particle at 
the position ${\bf x}_1$ at the time $t_1$, another
particle at the position ${\bf x}_2$ at the time $t_2$, etc.
(For a recent generalization of this to the relativistic case,
see \cite{nikolfol}.)
When different particles do not interact with each other, 
then the MFT wave function
satisfies $n$ independent local Schr\"odinger equations
\begin{equation}\label{schrl}
\hat{H}_i \Psi=i\frac{\partial}{\partial t_i} \Psi ,
\end{equation}
where
\begin{equation}
\hat{H}_i = -\frac{\nabla_i^2}{2m_i} +V_i({\bf x}_i,t_i) ,
\end{equation}
and $i=1,\ldots,n$. It is convenient to introduce a simpler notation
${\bf X}\equiv\{ {\bf x}_1,\ldots,{\bf x}_n \}$,
$T \equiv\{ t_1,\ldots,t_n \}$.
We also introduce global operators
\begin{equation}
\frac{\partial}{\partial T} = 
\sum_{j=1}^n \frac{\partial}{\partial t_j} , \;\;\;\;
\hat{H} = \sum_{j=1}^n \hat{H}_j .
\end{equation}
Thus, by summing up the local Schr\"odinger equations (\ref{schrl}), 
one obtains a single global Schr\"odinger equation
\begin{equation}\label{schr}
\hat{H} \Psi=i\frac{\partial}{\partial T} \Psi .
\end{equation}
The dynamics can be described by a Schr\"odinger equation of the 
form of (\ref{schr}) even when different particles do interact with each other.

The MFT Schr\"odinger equation (\ref{schr})
contains the ordinary
single-time Schr\"odinger equation as a special case
in which $t_1=\cdots =t_n\equiv t$. The corresponding wave functions are 
related as
\begin{equation}\label{pP}
\psi({\bf X},t)=\Psi({\bf X},t_1,\ldots,t_n)|_{t_1=\cdots =t_n= t} .
\end{equation}
However, the    
instantaneous synchronization in (\ref{pP}) is not more physical than,
for example, a relativistically more appealing
retarded light-cone synchronization. Indeed, the question of ``true"
synchronization in relativistic QM can be viewed as analogous
to the question of ``true" gauge in electrodynamics. In this analogy,
$\Psi({\bf X},T)$ is a ``gauge-independent" quantity, whereas
$\psi({\bf X},t)$ resembles the Coulomb gauge in which      
the electromagnetic potential propagates instantaneously.
(Of course, the analogy with gauge theories should not be
taken too literally, but note that a similar analogy with gauge
theories has been used in \cite{peres} as a response to the
criticism in \cite{nikpra}.)

A normalized solution $\Psi({\bf X},T)$ of (\ref{schr})
can be written as a linear combination of other orthonormal solutions
as
\begin{equation}\label{expan}
\Psi({\bf X},T)=\sum_a c_a \Psi_a({\bf X},T).
\end{equation}   
The base $\{ \Psi_a \}$ can be chosen such that each $\Psi_a$ is a 
local product of the form
\begin{equation}\label{locprod}
\Psi_a({\bf X},T) = \psi_{a1}({\bf x}_1,t_1) \cdots
\psi_{an}({\bf x}_n,t_n) .
\end{equation}
Thus, the base wave functions $\Psi_a({\bf X},T)$ do not exhibit
a nonlocal entanglement, but a general superposition (\ref{expan}) 
does.  

Now assume that $\psi_{a1}({\bf x}_1,t_1)$ are the
eigenstates of some local Hermitian operator that is measured.
Such a local measurement induces a {\em nonlocal} 
wave-function ``collapse" 
\begin{equation}\label{colaps}
\Psi({\bf X},T) \rightarrow \Psi_a({\bf X},T) .
\end{equation}
Now the crucial point is the following:
If the local measurement is performed at some particular value 
of the time $t_1$, then it does {\em not} 
mean that the {\em whole} wave function $\Psi({\bf X},T)$
collapses at the same particular value of time. Namely, fixing the 
value of $t_1$ in the collapsed wave function $\Psi_a({\bf X},T)$ 
in (\ref{colaps}) does {\em not} fix the values of $t_2,\ldots,t_n$.
In this sense, {\em in the MFT formulation of QM, the wave-function
``collapse" does not require any preferred notion of simultaneity.}
Thus the MFT formulation of QM can be used to enlighten 
the Einstein-Podolsky-Rosen effect (see e.g. \cite{ghose}) and the 
delayed-choice experiment (we are not aware of 
any particular reference that explicitly uses the MFT formulation to discuss 
the delayed-choice experiment).    

Concerning the problem of measurement, the only true problem in orthodox QM 
is to understand a physical mechanism that induces the 
wave-function ``collapse" (\ref{colaps}). Such a mechanism 
is provided by the MFT Bohmian hidden-variable interpretation studied 
in the next section.  

\section{MFT Bohmian interpretation of many-particle QM}
\label{BOHM}

By writing $\Psi=Re^{iS}$, where
$R$ and $S$ are real functions, the complex
equation (\ref{schr}) is equivalent to a set of two real equations
\begin{equation}\label{HJ}
\sum_{i=1}^n \left[ \frac{(\nabla_i S)^2}{2m_i}  
+V_i({\bf x}_i,t_i) \right] +Q({\bf X},T)
+\frac{\partial S}{\partial T} =0 ,
\end{equation}
\begin{equation}\label{eqvar}
\frac{\partial \rho}{\partial T} +
\sum_{i=1}^n \nabla_i \left( \rho \frac{\nabla_i S}{m_i} \right) =0 ,
\end{equation}
where $\rho=R^2$ and
\begin{equation}
Q=-\sum_{i=1}^n \frac{1}{2m_i} \frac{\nabla^2_i R}{R} .
\end{equation}
The conservation equation (\ref{eqvar}) confirms that
it is consistent to interpret $\rho({\bf X},T)$
as the probability density.

In analogy with the ordinary single-time Bohmian interpretation, 
we introduce a MFT beable ${\bf x}_i(T)$
that satisfies the MFT Bohmian equation of motion
\begin{equation}\label{bohmtraj}
\frac{\partial{\bf x}_i(T)}{\partial T} = \frac{\nabla_i S}{m_i} .
\end{equation}
From (\ref{bohmtraj}) and (\ref{HJ}), one can also derive the MFT 
quantum Newton equation
\begin{equation}
m_i\frac{d^2 {\bf x}_i(T)}{dT^2}=-\nabla_i [V_i({\bf x}_i,t_i) +
Q({\bf X},T) ] .
\end{equation}
In contrast with the ordinary Bohmian interpretation, 
the beable ${\bf x}_i(T)\equiv {\bf x}_i(t_1,\ldots,t_n)$ {\em cannot} be 
interpreted as a trajectory in spacetime. Nevertheless, for 
$t_1=\cdots =t_n=t$, the beable ${\bf x}_i(T)$ reduces to the 
ordinary Bohmian beable ${\bf x}_i(t)$, which, indeed, can be interpreted as 
a trajectory in spacetime. However, the fundamental ontology
is not represented by the synchronization-dependent function 
${\bf x}_i(t)$, but rather by the synchronization-independent function
${\bf x}_i(T)$. (Recall the analogy with gauge theories, discussed in the 
preceding section.)
For any set $T=\{ t_1,\ldots,t_n \}$, the functions 
${\bf x}_i(T)$, $i=1,\ldots,n$, uniquely specify the 
particle positions ${\bf x}_i$. Analogously to the 
ordinary Bohmian interpretation, Eqs.~(\ref{bohmtraj}) and 
(\ref{eqvar}) imply that the MFT Bohmian interpretation predicts 
the same probabilities for 
finding the first particle at
the position ${\bf x}_1$ at the time $t_1$, the second
particle at the position ${\bf x}_2$ at the time $t_2$, etc., 
as does the orthodox interpretation of MFT QM.
Moreover, if the wave functions $\Psi_a({\bf X},T)$ in (\ref{locprod}) 
do not overlap in at least a part of the configuration space, so that 
$\Psi_a({\bf X},T) \Psi_{a'}({\bf X},T)=0$ for $a\neq a'$, 
then some of the degrees of freedom can be interpreted as the degrees 
of freedom of the measuring apparatus. Consequently, analogously to the 
ordinary Bohmian interpretation, the MFT Bohmian interpretation 
predicts the same probabilites (equal to $|c_a|^2$)
for the effective ``collapse" (\ref{colaps}) 
as does the orthodox MFT interpretation. In the MFT Bohmian interpretation, 
the effective ``collapse" occurs because the beables ${\bf x}_i(T)$ 
take values from the support of one and only one of the nonoverlapping 
wave functions $\Psi_a({\bf X},T)$.  

Is the ontology represented by ${\bf x}_i(T)$ in contradiction with the fact 
that, for example, we can observe the particle position ${\bf x}_1$ at the time
$t_1$ {\em without} measuring the times $t_2,\ldots,t_n$? 
Although there is no beable corresponding to the quantity ${\bf x}_1$ at
time $t_1$, the beables ${\bf x}_i(T)$ determine the wave function 
$\Psi_a$ to which $\Psi$ will effectively ``collapse". If the functions 
$\Psi_a$ in (\ref{locprod}) are such that $\psi_{a1}({\bf x}_1,t_1)$ 
are eigenfunctions of the local position operator ${\bf x}_1$, then such 
a collapse can be viewed as a measurement of ${\bf x}_1(t_1)$, despite 
the fact that there is no beable corresponding to ${\bf x}_1(t_1)$.
Indeed, this is just an example of a measurement of an {\it un}preferred
observable in the Bohmian interpretation, such as momentum or 
energy in the ordinary single-time Bohmian interpretation. In the 
MFT Bohmian interpretation, the preferred observables are ${\bf x}_i(T)$, 
but the general theory of quantum measurements explains measurements 
of all other observables, with the same statistical predictions 
as in the orthodox interpretation. 

Let us also compare the nonlocality features in the ordinary 
and MFT Bohmian interpretations. In the ordinary 
single-time Bohmian interpretation, the ontology of hidden variables is 
classical at the kinematical level (given by local particle trajectories), 
whereas the quantum nonlocality is realized only on the 
dynamical level (encoded in the instantaneous nonlocal quantum potential).
In contrast, in the MFT Bohmian interpretation, the ontology 
is nonclassical and nonlocal already at the kinematical level, because, 
in ${\bf x}_i(T)$, ${\bf x}_i$ is a function not only of 
$t_i$, but of {\em all} $t_1,\ldots,t_n$. One may complain that 
the function ${\bf x}_i(T)$ is difficult to visualize, but one should 
not be worried about that, given the fact that it is certainly 
not more difficult to visualize than the MFT wave function 
$\Psi({\bf X},T)$. One should recall that, historically, the aim of 
the Bohmian interpretation was {\em not} to restore the classical ontology in 
QM (although, perhaps surprisingly, the ordinary 
Bohmian interpretation has done that), but rather to find {\em some} 
nonlocal beable that could reproduce the predictions of orthodox QM.   

We also note that
the MFT formalism enables one to formulate the Bohmian interpretation 
of many-particle systems in an explicitly relativistic-covariant way. 
This will 
be the subject of a separate paper, but we anticipate that it can be done 
by combining the results of the present paper with those of 
\cite{nikolfol}.

\section{MFT Bohmian interpretation of QFT}
\label{QFT}

The purpose of the present section is to generalize the results 
of the preceding sections to the case of QFT. However,
as the MFT Bohmian interpretation of QFT has already been discussed in
detail in \cite{nikolpla}, 
in this section we only briefly outline the main points of the
generalization, emphasizing those aspects that 
have been treated incorrectly in \cite{nikolpla},
or have not been discussed at all.

Instead with a discrete set ${\bf X}=\{ {\bf x}_1,\ldots,{\bf x}_n \}$, 
field theory deals with a continuous set of values of fields 
at different points, $\phi=\{ \phi({\bf x}) \}$, at all space points ${\bf x}$.
Similarly, the discrete set of times 
$T=\{ t_1,\ldots,t_n \}$ is replaced with a continuous set 
$T=\{ T(\bf x) \}$. The quantum state is represented by a 
wave functional $\Psi[\phi,T]$. The QFT analog of (\ref{schrl}) 
is known as the Tomonaga-Schwinger equation \cite{tomo,schw}.
Introducing the operator
\begin{equation}
\frac{\partial}{\partial T}=
\int d^3x'\, \frac{\delta}{\delta T({\bf x}')} ,
\end{equation}    
the QFT analog of the MFT Bohmian equation of motion (\ref{bohmtraj}) is 
\begin{equation}\label{bohmtrajqft}                         
\frac{\partial \phi({\bf x};T]}{\partial T} = 
\frac{\delta S}{\delta\phi({\bf x})} .
\end{equation}
(On the right-hand side, it is understood that 
$\phi({\bf x}')$ is replaced with $\phi({\bf x}';T]$ at all points ${\bf x}'$.) 
However, in \cite{nikolpla} it was stated that the fundamental 
MFT Bohmian equation was not the global MFT equation (\ref{bohmtrajqft}), 
but a local MFT equation
\begin{equation}\label{bohmtrajqftloc}                      
\frac{\delta \phi({\bf x};T]}{\delta T({\bf x}')} =
\delta^3 ({\bf x}-{\bf x}') 
\frac{\delta S}{\delta\phi({\bf x})} .
\end{equation} 
Indeed, if (\ref{bohmtrajqftloc}) is satisfied, then (\ref{bohmtrajqftloc})
implies (\ref{bohmtrajqft}). However, although Eq.~(\ref{bohmtrajqft})
is consistent, Eq.~(\ref{bohmtrajqftloc}), in general, may not be consistent.
In general, the right-hand side of (\ref{bohmtrajqftloc}) depends 
not only on $T({\bf x})$, but on the {\em whole} function $T$ at all 
points ${\bf x}'$. On the other hand, the $\delta$-function 
on the right-hand side of (\ref{bohmtrajqftloc}) implies that
$\phi({\bf x};T]$ on the left-hand side
does not depend on the whole function 
$T$, but only on $T({\bf x})$. However, for ${\bf x}'={\bf x}$,
this implies that the left-hand side 
of (\ref{bohmtrajqftloc}) depends only on $T({\bf x})$, whereas the 
right-hand side depends on the whole function $T$, which is inconsistent.
Thus, the correct MFT Bohmian equation of motion is 
(\ref{bohmtrajqft}), rather than (\ref{bohmtrajqftloc}).
Consequently,
contrary to the claim in \cite{nikolpla}, the MFT Bohmian beable 
is, in general, a genuine MFT field $\phi({\bf x};T]$, rather than a local 
field $\phi({\bf x},T({\bf x}))$. Nevertheless, the local appearance of
fields can be explained by the theory of quantum measurements, analogous 
to that in the preceding section.

It is also interesting to study the conditions under which the local 
MFT Bohmian equation of motion (\ref{bohmtrajqftloc}) could 
still be consistent.
One such condition is a wave functional that has a form of a local 
product analogous to (\ref{locprod}), but such a condition is not
sufficiently general. A more general condition is {\em any} 
quantum field theory that contains {\em gravity} as one of the quantized 
fields. Namely, the theories that contain gravity have a property 
of diffeomorphism invariance, which implies that the Hamiltonian always
vanishes on-shell. Consequently, instead of a functional Schr\"odinger
or Tomonaga-Schwinger 
equation, one deals with the Wheeler-DeWitt equation 
\cite{wh,dw,padm,kuc2,ish}
\begin{equation}
\hat{\cal H}({\bf x}) \Psi[g,\phi] =0 ,
\end{equation}  
where $\hat{\cal H}({\bf x})$ is the Hamiltonian-density operator,
$g$ represents the 3-metric and $\phi$ represents all other ``matter" 
fields. Since the wave functional $\Psi[g,\phi]$ does not depend on time
(either on $t$ or on $T$), 
it is consistent to postulate a local MFT Bohmian equation of motion of the 
form of (\ref{bohmtrajqftloc}) for both $\phi$ and $g$. 

Finally, let us note that it is straightforward to write all equations 
of this section in a manifestly general-covariant form, by using 
the formalism presented in \cite{nikolpla}. 
In particular, this leads to a covariant version
of the Bohmian interpretation of quantum gravity, which represents 
an improvement of the noncovariant Bohmian interpretation of quantum gravity
studied in \cite{holrep,gol,pin1,pin2,shoj}.

\section{Conclusion}
\label{CONCL}

The MFT formulation of QM and QFT allows a formulation of quantum theory 
that does not require a preferred definition of simultaneity, which 
alleviates the problem of relativistic-covariant formulation of 
quantum theory, including the problem of simultaneity of the 
wave-function ``collapse". The corresponding Bohmian interpretation
leads to new MFT beables that also do not require a preferred definition
of simultaneity. These MFT beables
have a manifest nonlocal nature already at the 
kinematical level. Nevertheless, the observed local appearance 
of particles and fields can be recovered by studying the theory of quantum
measurements.   

\section*{Acknowledgements}
This work was supported by the Ministry of Science and Technology of the
Republic of Croatia. 


\begin{thebibliography}{99}
\bibitem{medina}
R. Medina, Annales de la Fondation Louis de Broglie 24 (1999) 129,
quant-ph/0508014.
\bibitem{bell}
J.S. Bell, Speakable and Unspeakable in Quantum Mechanics, 
Cambridge University Press, Cambridge, 1987.
\bibitem{bohm12}
D.~Bohm, Phys.~Rev.~85 (1952) 166, 180.
\bibitem{bohmrep1}
D.~Bohm, B.J.~Hiley, 
Phys.~Rep.~144 (1987) 323.
\bibitem{bohmrep2}
D.~Bohm, B.J.~Hiley, P.N.~Kaloyerou,
Phys.~Rep.~144 (1987) 349.
\bibitem{holrep}
P.R.~Holland, Phys.~Rep.~224 (1993) 95.
\bibitem{holbook}
P.R.~Holland, The Quantum Theory of Motion,
Cambridge University Press, Cambridge, 1993.
\bibitem{nikfpl12}
H.~Nikoli\'c, Found.~Phys.~Lett.~17 (2004) 363;
H.~Nikoli\'c, Found.~Phys.~Lett.~18 (2005) 123.
\bibitem{durr99}
D.~D\"urr, S.~Goldstein, K.~M\"unch-Berndl, N.~Zangh\`i,
Phys.~Rev.~A 60 (1999) 2729.
\bibitem{hort}
G.~Horton, C.~Dewdney, J.~Phys.~A 37 (2004) 11935.
\bibitem{nikolDDW}
H.~Nikoli\'c, Eur. Phys. J. C 42 (2005) 365;
H.~Nikoli\'c, hep-th/0512186; 
H.~Nikoli\'c, hep-th/0601027.
\bibitem{durr96}
K.~Berndl, D.~D\"urr, S.~Goldstein, N.~Zangh\`i,
Phys.~Rev.~A 53 (1996) 2062.
\bibitem{nikolcov}
H.~Nikoli\'c, Found. Phys. Lett. 18 (2005) 549;
H.~Nikoli\'c, quant-ph/0512065.
\bibitem{nikolpla}
H.~Nikoli\'c, Phys. Lett. A 348 (2006) 166.
\bibitem{tomo}
S.~Tomonaga, Prog.~Theor.~Phys.~1 (1946) 27.
\bibitem{schw}
J.~Schwinger, Phys.~Rev.~74 (1948) 1439.
\bibitem{nikolfol}
H.~Nikoli\'c, quant-ph/0602024.
\bibitem{peres}
A. Peres, Phys. Rev. A 64 (2001) 066102.
\bibitem{nikpra}
H.~Nikoli\'c, Phys. Rev. A 64 (2001) 066101.
\bibitem{ghose}
P.~Ghose, D.~Home, Phys.~Rev.~A 43 (1991) 6382.
%
\bibitem{wh}  
J.A.~Wheeler, in Battelle Rencontres, eds. C.M.~DeWitt,
J.A.~Wheeler, Benjamin, New York, 1968.
\bibitem{dw}  
B.~DeWitt, Phys.~Rev. 160 (1967) 1195.
\bibitem{padm}
T.~Padmanabhan, Int.~J.~Mod.~Phys. A4 (1989) 4735.
\bibitem{kuc2}
K.~Kucha\v r, in Proceedings of the 4th Canadian Conference
on General Relativity and Relativistic Astrophysics,
World Scientific, Singapore, 1992.
\bibitem{ish} 
C.J.~Isham, gr-qc/9210011.
%
\bibitem{gol} 
S.~Goldstein, S.~Teufel, quant-ph/9902018.
\bibitem{pin1}
N.~Pinto-Neto, E.S.~Santini, Phys.~Rev.~D 59 (1999) 123517.
\bibitem{pin2}
N.~Pinto-Neto, E.S.~Santini, Gen.~Rel.~Grav.~34 (2002) 505.
\bibitem{shoj}
A.~Shojai, F.~Shojai, Class.~Quant.~Grav.~21 (2004) 1.


\end{thebibliography}
\end{document}